\documentclass[12pt]{iopart}

\usepackage{iopams}  
\usepackage{subcaption}
\usepackage{graphicx}
\usepackage{siunitx}
\usepackage{comment}

\begin{document}

\title[Temperature dependent multi-pulse threshold due to SWCNT/PDMS Saturable Absorber]{Temperature dependent multi-pulse threshold due to SWCNT/PDMS Saturable Absorber}

\author{Rhona L Hamilton,  Moeri Horiuchi, Bowen Liu, Takuma Shirahata, Sze Y Set and Shinji Yamashita}

\address{Research Center for Advanced Science and Technology, The University of Tokyo, 4-6-1 Komaba, Meguro-ku Tokyo 153-8904, Japan}
\ead{hamilton-rhona@g.ecc.u-tokyo.ac.jp}
\vspace{10pt}
\begin{indented}
\item[]May 2024
\end{indented}

\begin{abstract}
The threshold pump power for modelocking decreased by 18\% when the temperature was increased from $25\si{\degreeCelsius}$ to $100\si{\degreeCelsius}$ where a SWCNT/PDMS coated tapered fiber was used as the saturable absorber in a fiber laser. Further, the pump power at which multi-pulse operation began decreased by 24\%, and the pump power range over which fundamental modelocking could be maintained decreased by 59\% over the same temperature range. This decrease in stability is attributed to the large thermo-optic coefficient of the PDMS polymer, which results in a 40\% reduction of the overlap between the evanescent field and SWCNT coating of the taper fiber over a temperature range of $75\si{\degreeCelsius}$.
\end{abstract}

%
%
%
%
%

\section{Introduction}
Fiber lasers, which can be made to produce femtosecond duration pulses through modelocking have an immense range of applications, including material processing \cite{Liu_1997}, supercontinuum generation \cite{Birks:00}, and frequency comb generation for precision frequency metrology \cite{Holzwarth_2001}. 
Single walled carbon nanotubes (SWCNTs) have been widely researched as a saturable absorber (SA), due to their broadband absorption, high damage threshold, and ultra-fast recovery times \cite{Yamashita:12}. Further, SWCNTs can also be incorporated into an all-fiber laser configuration in a number of ways. One such technique is to coat a fiber ferrule with a layer of SWCNTs \cite{Set_2004}. While advantageous in its simplicity, this method has the disadvantage that the laser cavity powers are limited to avoid causing damage to the SWCNT SA. An alternative method is to apply the SWCNTs to a taper fiber and then incorporate this all fiber device into the laser cavity. 

A taper fiber, or microfiber, refers to an optical fiber with one section of the fiber significantly reduced in diameter compared to the rest of the fiber \cite{Birks_and_Li_92}. This region is referred to as the taper waist, and typically has a diameter on the order of a few microns. The result of tapering is that the fundamental mode of the fiber, which is ordinarily confined to core, becomes a cladding mode in the taper waist. Here, the light can interact with the surrounding environment through evanescent field coupling. This has the advantage that there is no optically induced damage to the SA, even for high circulating pulse powers. The incredibly narrow diameter of the taper waist increases the fragility of the device and makes it more environmentally sensitive compared to non-tapered fiber. 

\begin{figure}[h]
    \centering
    \begin{minipage}{0.9\textwidth}
        \centering
        \includegraphics[width=1\textwidth]{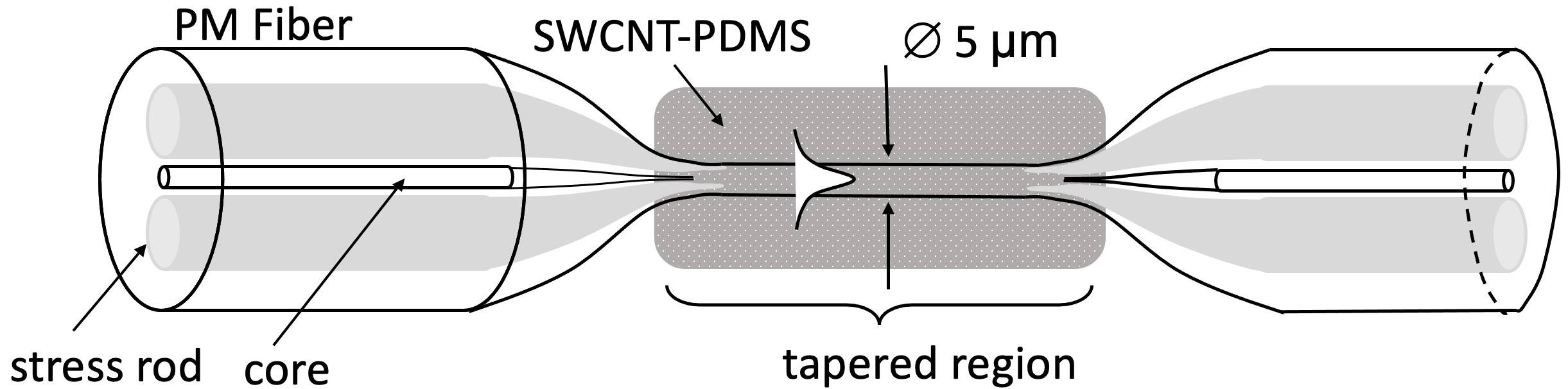}
        \caption{ (a) PM tapered fiber device with SWCNT/PDMS composite.  }
        \label{fig:tapered_fiber}
    \end{minipage}
    \end{figure}

There are several methods of applying the SWCNTs to the taper fiber. These include spray coating \cite{Song:2007} and optical deposition \cite{Kashiwagi:09} . The SWCNTs can also be applied to the taper fiber in solution with a polymer, which has the advantage that the polymer coating can make the tapered area more robust and protect it from environmental variations \cite{Kieu:07}. In all of the above methods an important consideration is to minimise agglomeration of the SWCNTs into micron scale structures, which otherwise results in excess loss due to scattering \cite{Ma_jen_dalton}.

\begin{table*}[t]
\begin{center}
\begin{tabular}{ccccc} 
Polymer & n (room temp)  & dn/dT  [/$^o$C]      & Water soluble & Tg [$^o$C]    \\
\hline
PMMA    & 1.49   \cite{Ma_jen_dalton} & -1.3   x 10-4 \cite{ZHANG2006} & no & 105  \cite{Hajduk2021} \\
PC      & 1.58   \cite{Ma_jen_dalton} & -0.9   x 10-4 \cite{ZHANG2006} & no& 143 \cite{Sehrawat2022} \\ 
PS      & 1.59   \cite{Ma_jen_dalton} & -1.2   x 10-4 \cite{ZHANG2006} & no  & 107  \cite{Rieger1996}  \\
PDMS    & 1.41  \cite{Zhu:17} & -4.5   x 10-4 \cite{Zhu:17} & no  & -135  \cite{KLONOS2018169} \\
PVA     & 1.51  \cite{Gad2023} & -1.2   x 10-4 \cite{BENDOUDOU2021} & yes   & 85 \cite{THOMAS2018145}
\end{tabular}
\caption{Physical parameters of common polymers. PMMA - Poly(methyl methacrylate), PC - Polycarbonate, PS - Polystyrene, PDMS - Polydimethylsiloxane, PVA - Polyvinyl Alcohol. Parameters, n - refractive index, dn/dT - thermo-optic coefficient, water solubility, and T$_g$ - glass transition temperature, are chosen to quantify environmental sensitivity and transparency of the polymers.}
\label{tab:polymer}
\end{center}
\end{table*}

There has been much research conducted in fabricating SWCNT/Polymer composites, with a wide variety of polymers used. Table \ref{tab:polymer} lists a selection of the most commonly available and widely studied polymers for this application, along with their relevant physical properties. 

There are three key parameters in deciding whether a polymer is suitable for using in a composite with SWCNT applied to a taper fiber. These are environmental stability, transparency, and the ease with which SWCNTs can be dispersed through the polymer matrix \cite{Tawfique_2009}. 
 
The environmental stability of a polymer is determined by the sensitivity of the polymer to thermal effects, water and humidity. Additionally, the polymer must be resistant to not only heating from the environment but also that induced by the laser operation. Table \ref{tab:polymer} lists the thermo-optic coefficient (dn/dT), the water solubility, and the glass transition temperature (T$_g$) in an attempt to quantify environmental stability to some degree.  

Furthermore, the polymer should have low absorption losses at the lasing wavelength. At telecoms wavelengths, these losses are due to higher harmonics of molecular vibration resonances \cite{Ma_jen_dalton}. In the long term, the transparency of polymers tends to decrease due to oxidation and the expulsion of H-halogen molecules \cite{Tawfique_2009}. On the time scale of the laser's operation one must also consider the change in transparency arising from thermo-optic effects.  

To minimise scattering loss the SWCNTs must be uniformly dispersed in the polymer matrix. This places further requirements on the polymers used. Pristine SWCNTs do not disperse easily. One method of increasing SWCNT dispersion is called covalent functionalisation, where functional groups are attached at defects in the SWCNT side walls \cite{Hirsh_2002}. However, the optical properties of SWCNTs are easily changed by this functionalisation \cite{Bahr2001} and it is therefore not preferentially used in photonics applications. Instead, solvents and dispersants are used which interact non-covalently with the SWCNT sidewalls, minimally changing the optical and electronic properties \cite{Britz_2006}. Non-water soluble polymers (PMMA, PC, PS, PDMS) require non-aqueous solvents to form well dispersed composites with SWCNTs. On the other hand, while the environmental stability of PVA is lower due to its water solubility, high quality SWCNT composites are easily prepared using aqueous solvents \cite{SCARDACI2007}. 

In this work we present the temperature dependent loss of a SWCNT/PDMS taper fiber SA. We also investigate the impact this has on the operation of a fiber laser modelocked with the SA, and demonstrate that the pump range for stable fundamental modelocking decreases by 59\% when the temperature is increased by 75$^o$C. A mechanism for this decrease in stability with increasing temperature is also proposed. 
 
\section{Temperature dependent loss measurement}

A single mode polarisation maintaining (PM) fiber was tapered by controlled stretching of the fiber in a heater. The fiber was heated to a temperature of 1600$^o$C, at which the glass can be deformed. The fiber was held between two translation stages which were moved apart at an equal, steady rate to ensure equal force on both ends of the fiber.  During the tapering process the insertion loss of the fiber was measured to monitor the condition of the taper. One end of the fiber was connected to an erbium doped fiber amplifier (EDFA), which served as a light source, and the other end was connected to a power meter \cite{Kashiwagi:09}.  

The SWCNT/PDMS composite was prepared by sonicating SWCNTs (commercial HiPCO SWCNT 1\% wt. concentration) in solution with isopropyl alcohol (IPA) before mixing with SYGARD-184 PDMS resin (Dow Corning) through further sonication. This was then magnetically stirred until the IPA evaporated, after which the PDMS curing agent was added. The purpose of the sonication and magnetic stirring was to minimize the loss due to SWCNT aglomeration. The composite was applied to the tapered fiber, as in Fig. 1 (a). The complete SWCNT/PDMS fiber device was left to cure overnight, kept at a constant temperature of 60$^o$C.

\begin{figure}[h]
    \centering
    \begin{minipage}{0.90\textwidth}
        \centering
        \includegraphics[width=1\textwidth]{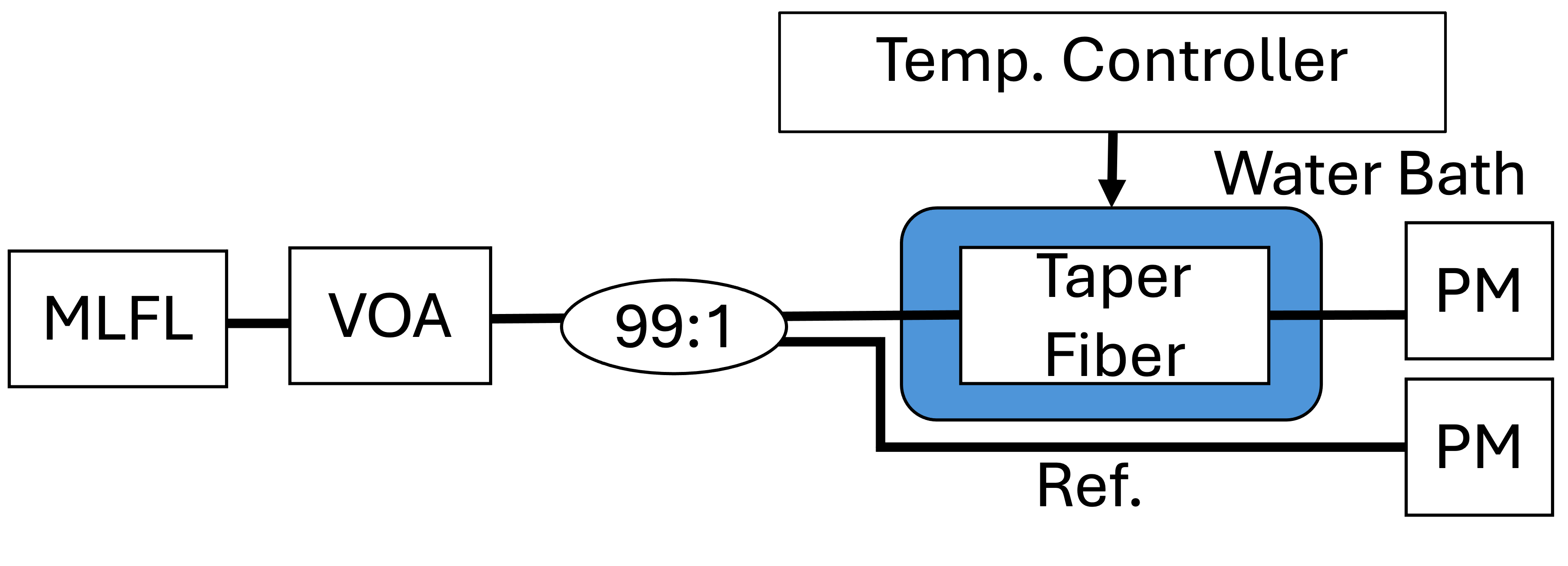}
        \caption{Set-up used to measure the saturable absorption of SWCNT/PDMS coated tapered fiber at varying temperatures. MLFL - modelocked fiber laser, VOA - variable optical attenuator. }
        \label{fig:waterBath}
    \end{minipage}
    \end{figure}
 
\begin{figure}[h]
    \centering
    \begin{minipage}{0.90\textwidth}
        \centering
        \includegraphics[width=1\textwidth]{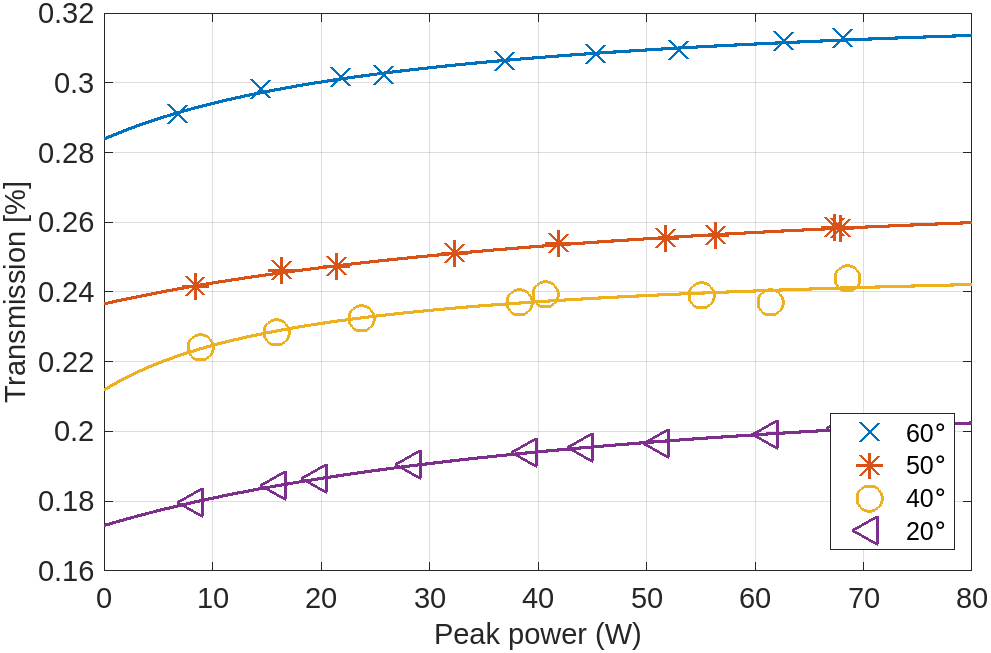}
        \caption{Transmission through SWCNT/PDMS coated tapered fiber as a function of peak power.}
        \label{fig:nonLin_T}
    \end{minipage}
    \end{figure}

An experiment was conducted to test the temperature dependence of the transmission through the SWCNT/PDMS taper fiber. A schematic of the experimental set up is given in Figure \ref{fig:waterBath}. The temperature of the taper fiber was varied by immersing it in a temperature controlled water bath. The source of input light to the system was a commercial modelocked fiber laser (MLFL) (Alnair Labs PFL-100) with a 1560 nm central wavelength, 6 dBm output power, 820 fs pulse width and 50 MHz repetition rate. The input power to the taper fiber was controlled by a variable optical attenuator (VOA). 1\% of this light was tapped to estimate the total input power. The output power from the taper fiber was measured by a second power meter. These measurements were done with the water at room temperature, taken to be $20\si{\degreeCelsius}$, and then repeated with the water heated to $40\si{\degreeCelsius}$, $50\si{\degreeCelsius}$ and $60\si{\degreeCelsius}$. 
The transmission through the SWCNT/PDMS taper fiber was calculated as a function of the MLFL peak power. This data was fitted to the power dependent loss $\alpha(P)$ for an instantaneous SA \cite{Haus1975}, given by 

\begin{equation}
\centering
 \alpha(P) = \frac{\alpha_0}{1+P/P_{\mathrm{sat}}} + \alpha_{ns},   
\end{equation}

 where $\alpha_0$ is the saturable loss parameter, $P_{\mathrm{sat}}$ is the saturation power and $\alpha_{ns}$ is the non-saturable loss. Both the data and fit are presented in Figure \ref{fig:nonLin_T}. The fitted parameters, in addition to the calculated modulation depth (MD) are given in Table \ref{tab:fitted_params}. The modulation depth is calculated as the percentage change in absorption from the non-saturated to the saturated state as given by the fitted loss parameters. 

Both $\alpha_{ns}$ and $\alpha_{0}$ decrease with increasing temperature. $\alpha_{0}$ decreases by 48\% and $\alpha_{ns}$ decreases by 25\% as the temperature increases from $20\si{\degreeCelsius}$ to $60\si{\degreeCelsius}$. Neither the saturation power nor the modulation depth show a clear trend in their change as a function of temperature. 

\begin{table}[]
\centering
\begin{tabular}{lllll}
Temp.[$^o$C] & $\alpha_0$ [m$^-1$] & P$_{\mathrm{sat}}$ & $\alpha_{\mathrm{ns}}$[m$^-1$]& MD [\%] \\
\hline
20                  & 5.05                                & 21.0    & 30.0                                   & 4.97           \\
40                  & 3.34                                & 8.4   & 27.7                                 & 3.84         \\
50                  & 3.17                                & 27.4  & 25.7                                 & 4.07         \\
60                  & 2.72                                & 15.5  & 22.5                                 & 4.13        
\end{tabular}
\caption{Fitted parameters from Equation 1, for temperature dependent loss of SWCNT/PDMS taper fiber SA. MD - modulation depth.}
\label{tab:fitted_params}
\end{table}

\section{Temperature dependent fundamental modelocking range}
After establishing the temperature dependence of the saturable and non-saturable loss in the SWCNT/PDMS SA, a further experiment was conducted to determine what effect this temperature dependence had on modelocking when the SA was incorporated into a fiber laser. 

\begin{figure}[h]
\centering
    \begin{minipage}{0.90\textwidth}
        
        \includegraphics[width=1\textwidth]{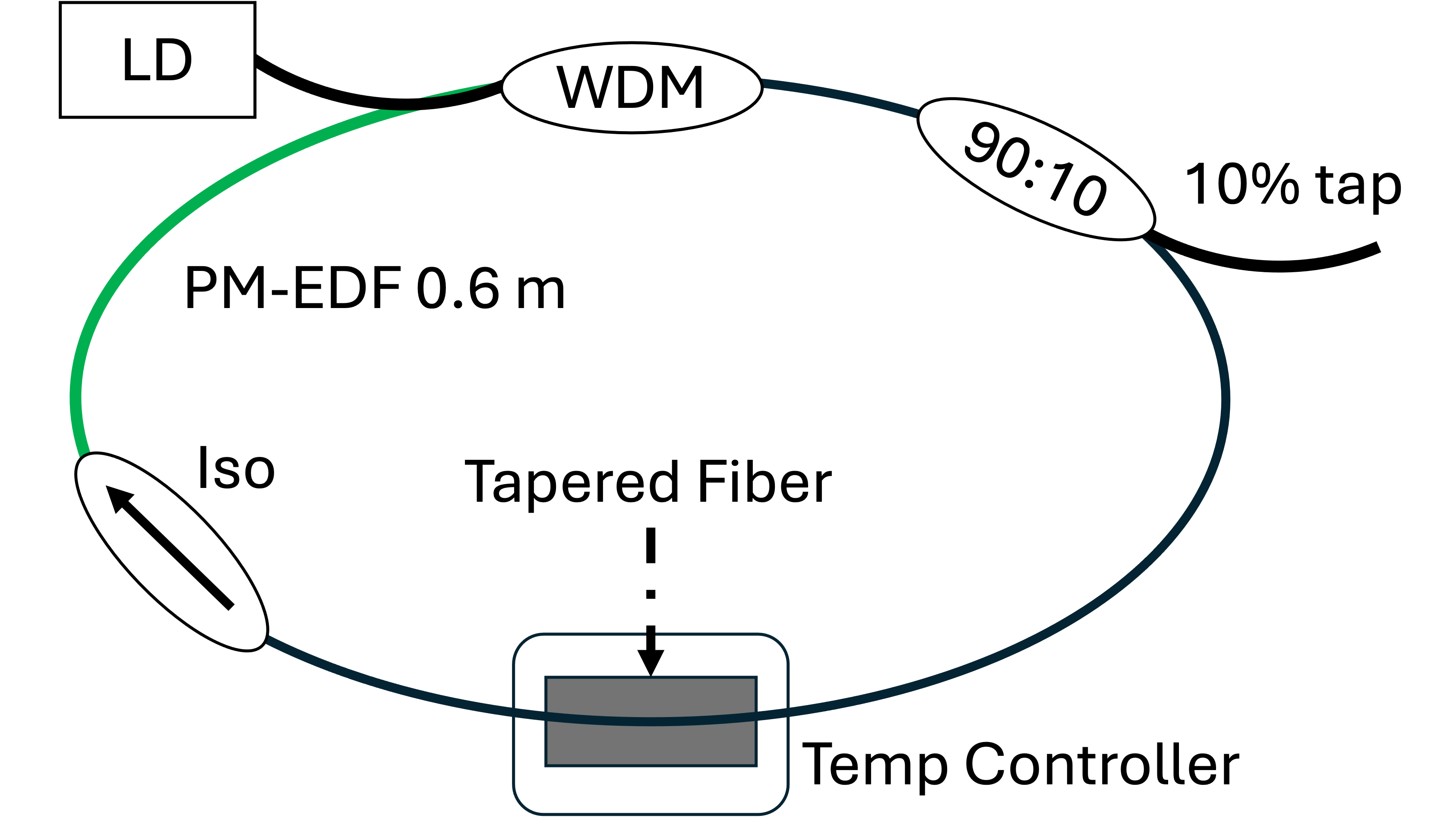}
        \caption{Set-up of mode-locked fiber laser with SWCNT/PDMS saturable absorber. LD - laser diode, WDM - wavelength division multiplexer, Iso - isolator, EDF - erbium doped fiber. }
        \label{fig:RingLaser}
    \end{minipage}
\end{figure}

The configuration of the fiber laser used in this experiment is given in Figure \ref{fig:RingLaser}. The pump light source was a single-mode 980 nm laser diode, which was coupled into the ring laser geometry through a 980/1550 WDM. 0.6 m of polarisation maintaining erbium doped fiber (PM-EDF) were used as the gain medium and 10\% of the circulating power was tapped at the output coupler. An isolator was included between the SWCNT/PDMS taper fiber and the PM-EDF to ensure unidirectional operation. The SWCNT/PDMS taper fiber was placed on a heater to vary the temperature during the experiment. All devices and fiber used in the laser were polarisation maintaining. 

\begin{figure}[h]
\centering
    \begin{minipage}{0.90\textwidth}
        
        \includegraphics[width=1\textwidth]{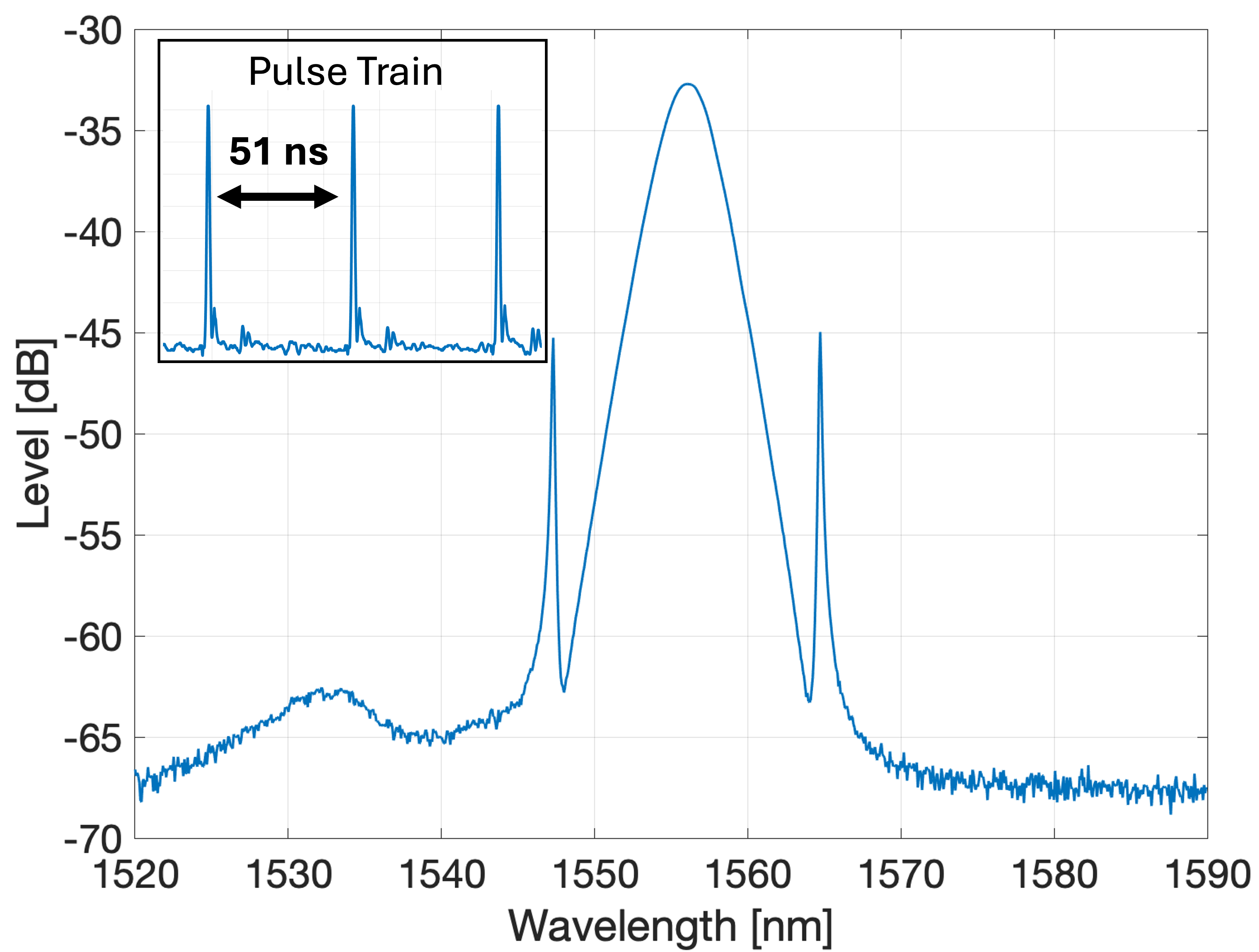}
        \caption{Fundamental modelocking spectrum at room temperature. Inset: pulse train. }
        \label{fig:spectrum}
    \end{minipage}
\end{figure}

At room temperature ($25^{\circ}\mathrm{C}$) fundamental modelocking was achieved at a minimum pump power of 47 mW. The laser was modelocked with a repetition rate of 19.6 MHz The optical spectrum and the pulse train are given in Figure \ref{fig:spectrum}. The 3dB bandwidth of the spectrum was measured to be 3.5 nm. The pulse duration was measured with a FROG autocorelator, giving a FWHM of 1.41 ps, and hence a pulse duration of 916 fs, assuming a sech$^2$ pulse. For pump powers greater than 57 mW, splitting of the soliton into multiple pulses was observed. This was taken to be the upper limit on the range of fundamental modelocking. 

\begin{figure}[h]
    \centering
    \begin{minipage}{0.90\textwidth}
        
        \includegraphics[width=1\textwidth]{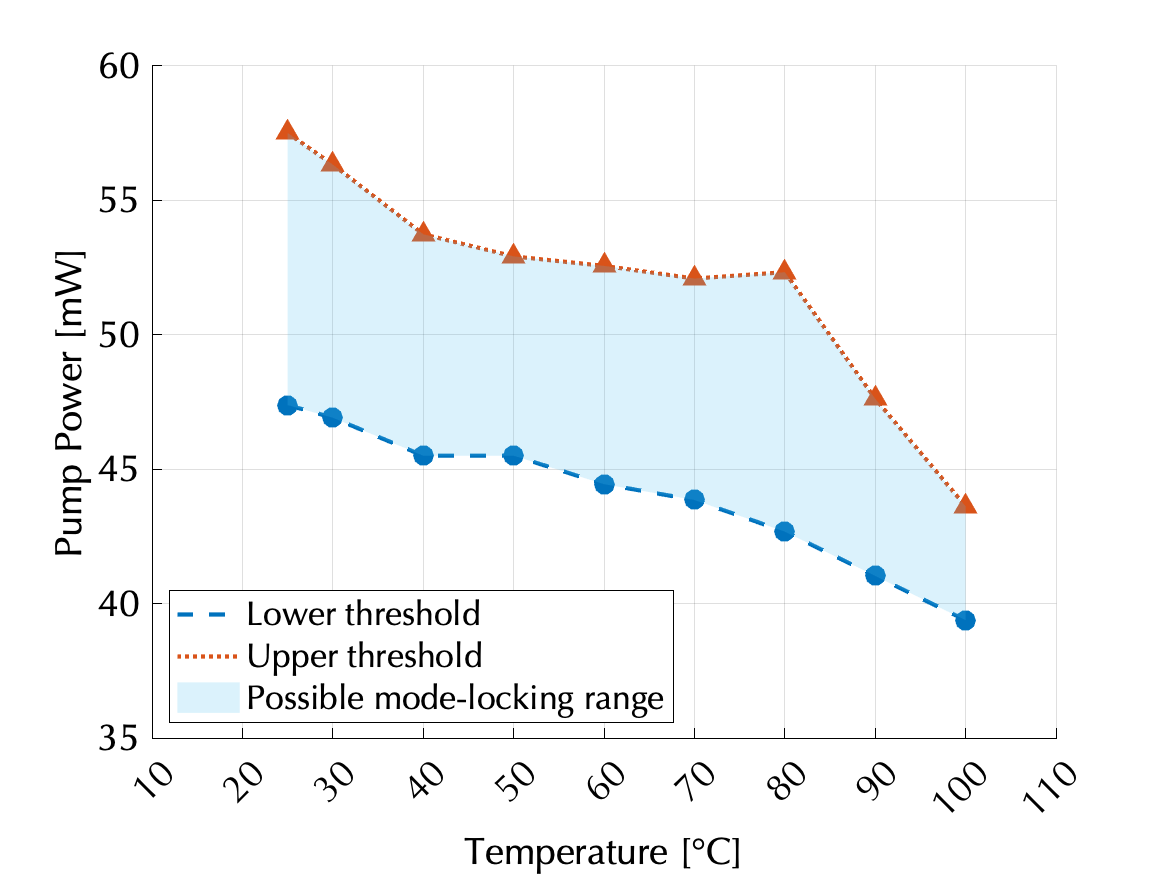}
        \caption{Possible range of pump power for fundamental modelocking. }
        \label{fig:ML_range}
    \end{minipage}

\end{figure}

This measurement of the fundamental modelocking range was repeated at higher temperatures. The SWCNT/PDMS SA was heated from $25^{\circ}\mathrm{C}$ to $100^{\circ}\mathrm{C}$ in $5^{\circ}\mathrm{C}$ increments. At each temperature the LD current was increased from threshold, noting when fundamental modelocking began and ceased. The results are presented in Figure \ref{fig:ML_range}.

\section{Discussion}
This change in the loss of the SWCNT/PDMS taper fiber SA is attributed to the large thermo-optic coefficient of PDMS, resulting in a significant change in refractive index over the range of temperatures considered. 
At higher temperatures the refractive index of the PDMS drops, resulting in a greater refractive index difference between the taper fiber cladding and PDMS coating, and hence a greater degree of confinement of the light propagating in the taper waist. This increased mode confinement in turn results in a decrease in the overlap between the evanescent field and the SWCNT/PDMS coating, and hence the light experiences less loss at higher temperatures. 

We attribute the decrease $\alpha_0$, $\alpha_{ns}$, and threshold pump power to a reduction in the evanescent field-SWCNT/PDMS overlap, resulting from the change in $n_{\mathrm{PDMS}}$ with temperature. To support this, we estimate the change in confinement of light with the change in ($\mathrm{n}_{\mathrm{PDMS}}$), assuming the refractive index of SWCNT/PDMS to be $\mathrm{n}_{\mathrm{PDMS}}$ as PDMS makes up 99\% of the compound. The normalised wavelength, V, is given by $V = (2\pi/\lambda)a\sqrt{n_1^2 - n_2^2},$ where $\lambda$ is the wavelength of light in the fiber, \textit{a} is the taper waist radius, and here $n_1$ and $n_2$ are $n_{\mathrm{cladding}}$ and $n_{\mathrm{PDMS}}$ respectively. We obtain $V(25^{\mathrm{o}}C) = 4.1$ and $V(100^{\mathrm{o}}C) = 5.1$, which, using the relation in \cite{OKAMOTO200657}, corresponds to a 5\% and 3\% fraction of the total power in the evanescent field respectively, or a 40\% reduction in the evanescent field-SWCNT/PDMS overlap. 

While the SWCNT/PDMS coated taper fiber has many advantages, this result introduces an additional environmental sensitivity of the device, namely, the temperature sensitivity of the PDMS refractive index. The refractive index of the polymer can be seen as a degree of freedom in the laser design, providing an opportunity to actively modulate or control the refractive index, and hence the loss, through the use of an electro-optic polymer. 

\section{References}
\bibliographystyle{unsrt}
\bibliography{TempSens}

\end{document}